\documentclass[twocolumn,aps,showpacs,superscriptaddress,floatfix]{revtex4}
\usepackage{graphicx}

\newcommand{\bnmr}{$\beta$-NMR}
\newcommand{\NS}{NbSe$_2$}
\newcommand{\Li}{$^8$Li}
\newcommand{\msr}{$\mu$SR}

\begin{document}

\title{Giant Vortices Below the Surface of NbSe$_2$ Detected Using Low Energy $\beta$-NMR}
\author{Z. Salman}
\affiliation{TRIUMF, 4004 Wesbrook Mall, Vancouver, BC, Canada, V6T 2A3}
\affiliation{Clarendon Laboratory, Department of Physics, Oxford
  University, Parks Road, Oxford OX1 3PU, UK}
\author{D.~Wang}
\affiliation{Department of Physics and Astronomy, University of
  British Columbia, Vancouver, BC, Canada V6T 1Z1}
\author{K.~H.~Chow}
\affiliation{Department of Physics, University of Alberta, Edmonton,
  AB, Canada T6G 2J1}
\author{M.D.~Hossain}
\affiliation{Department of Physics and Astronomy, University of
  British Columbia, Vancouver, BC, Canada V6T 1Z1} 
\author{S.R. Kreitzman} 
\affiliation{TRIUMF, 4004 Wesbrook Mall, Vancouver, BC, Canada, V6T 2A3}
\author{T.A.~Keeler}
\affiliation{Department of Physics and Astronomy, University of
  British Columbia, Vancouver, BC, Canada V6T 1Z1}
\author{C.D.P. Levy} 
\affiliation{TRIUMF, 4004 Wesbrook Mall, Vancouver, BC, Canada, V6T 2A3}
\author{W.A.~MacFarlane}
\affiliation{Chemistry Department, University of British Columbia,
  Vancouver, BC, Canada V6T 1Z1}
\author{R.I.~Miller}
\author{G.D.~Morris}  
\affiliation{TRIUMF, 4004 Wesbrook Mall, Vancouver, BC, Canada, V6T 2A3}
\author{T.J.~Parolin}
\affiliation{Chemistry Department, University of British Columbia,
  Vancouver, BC, Canada V6T 1Z1}
\author{H.~Saadaoui}
\affiliation{Department of Physics and Astronomy, University of
  British Columbia, Vancouver, BC, Canada V6T 1Z1}
\author{M.~Smadella}
\affiliation{Department of Physics and Astronomy, University of
  British Columbia, Vancouver, BC, Canada V6T 1Z1}
\author{R.F. Kiefl}
\affiliation{TRIUMF, 4004 Wesbrook Mall, Vancouver, BC, Canada, V6T 2A3}
\affiliation{Department of Physics and Astronomy, University of
  British Columbia, Vancouver, BC, Canada V6T 1Z1}
\affiliation{Canadian Institute for Advanced Research, Canada}

\begin{abstract}
A low energy radioactive beam of polarized \Li\ has been used to
observe the vortex lattice near the surface of superconducting
\NS. The inhomogeneous magnetic field distribution associated with the
vortex lattice was measured using depth-resolved $\beta$-detected
NMR. Below $T_c$ one observes the characteristic lineshape for a
triangular vortex lattice which depends on the magnetic penetration
depth and vortex core radius. The size of the vortex core varies
strongly with magnetic field. In particular in a low field of $10.8$
mT the core radius is much larger than the coherence length. The
possible origin of these giant vortices is discussed.
\end{abstract}
\maketitle

The vortex state of a superconductor exhibits many fascinating
properties. A fundamental feature, which follows from the topology of
the superconducting ground state and electron pairing, is that each
vortex carries an elementary quantum of magnetic flux
$\phi_0=hc/2e$. In the simple Ginzburg-Landau model the order
parameter for superconductivity is zero at the vortex core and rises
to the bulk value on the scale of the coherence length, while the
magnetic field falls away from the core on the scale of the London
penetration depth. However, even in conventional superconductors
vortices are complex objects with properties that were not anticipated
when they were first predicted by Abrikosov \cite{Abrikosov57}. For
example, Caroli {\it et~al.} proposed the existence of bound
quasi-particles associated with an isolated vortex \cite{Degennes64}
which were eventually observed with STM measurement on \NS\
\cite{Hess89}. Thermal deoccupation of these bound states leads to
shrinking of the vortex core at low temperatures or Kramer-Pesch effect \cite{Kramer74}. Muon
spin rotation (\msr ) has been used extensively to measure the
magnetic field distribution in the vortex state which is a sensitive
probe of the vortex properties. For example \msr\ results on \NS\
\cite{Miller00} have shown that the low temperature core radius is
significantly larger than predicted in the quantum limit
\cite{Hayashi98}. Recently there has been considerable work on the
role of delocalized quasiparticles and the interaction between
vortices, particularly in multiband superconductors such as \NS\ and
MgB$_2$, where there is more than one superconducting gap. It is
believed such effects are responsible for the unusually large and
field dependent specific heat \cite{Sonier99} and thermal conductivity
\cite{Boaknin03} in the vortex state.

In this letter we report observation of giant vortices near the
surface of superconducting \NS\ using a novel method of low energy
$\beta$-detected NMR (\bnmr) \cite{Morris04, Salman06}. The technique
is similar in principle to \msr\ and in particular low energy \msr\
\cite{Morenzoni03}, but provides complementary information
\cite{Kiefl03}. Below $T_c$, we observe a broad asymmetric \bnmr\
lineshape, which is characteristic of a triangular lattice of magnetic
vortices. As in the case of \msr\ the internal field distribution, as
reflected by the lineshape, is sensitive to the magnetic penetration
depth and vortex core radius. Surprisingly the fitted core radius of
$77(10)$ nm is much larger than the coherence length ($\sim 10$ nm) in a
small magnetic field of $10.8$ mT. We propose the extended nature of
the vortices originates from multiband effects and thermal vibrations
of the vortices.

The low energy ($30$ keV) beam of \Li\ is produced at the isotope
separator and accelerator (ISAC) at TRIUMF. A large nuclear
polarization (70\%) is generated inflight using a collinear optical
pumping method. The \bnmr\ spectrometer sits on a high voltage
platform so that the implantation energy can be varied between
$1-30$~keV, corresponding to an average implantation depth between
$5-150$~nm. In \bnmr\ the nuclear polarization is monitored through
the anisotropic $\beta$ decay. In the case of \Li, which has a mean
lifetime of 1.2 s, the emitted betas have an average energy of about
$6$ MeV, so that they easily pass through stainless steel windows in
the ultra high vacuum (UHV) chamber. The magnetic resonance is
detected by monitoring the time-averaged nuclear polarization as a
function of a small perpendicular radio frequency magnetic field. \Li\
is a spin $I=2$ nucleus with a small electric quadrupole moment
$Q=+33$~mB and gyromagnetic ratio $\gamma=6.301$ MHz/T. In the absence
of a quadrupolar splitting the lineshape is determined by the
distribution of local magnetic fields at the Li site.

The single crystal of 2H-\NS, measuring about $4$ mm in diameter and
$0.1$ mm thick, was attached to a sapphire plate and mounted on a cold
finger cryostat. It had a sharp superconducting transition at
$T_c=7.0$ K with a width of $0.1$ K. The sample was cleaved just prior
to introduction into the UHV ($10^{-9}$ torr); such short exposure to
air does has no effect on our \bnmr\ measurements. The beam was
focused onto the sample so that there was no detectable background
signal from the sapphire.

\begin{figure}[h]
\includegraphics[width=8.0cm]{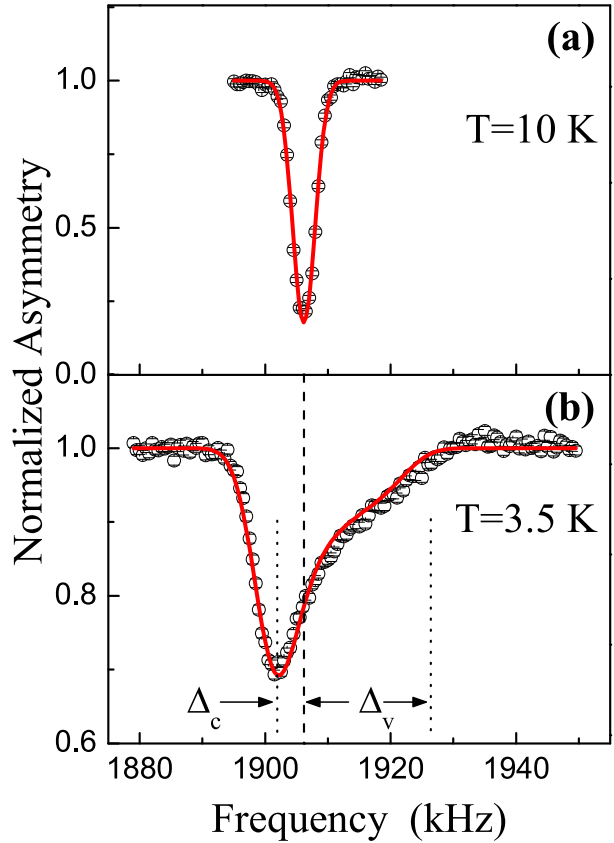}
\caption{(Color online) (a)The \bnmr\ spectrum in the normal state of
\NS\ at $10$ K in a magnetic field of $300$ mT applied along the
c-axis. The beam energy of $20$ keV corresponds to a mean implantation
depth $\langle z \rangle =85$ nm. The observed Gaussian line
broadening is attributed to $^{93}$Nb nuclear moments. (b) The same
conditions as (a) except field cooled to $3.5$ K or $0.5 T_c$. The
asymmetric lineshape is characteristic of a triangular lattice of
magnetic vortices.}
\label{highfieldlines}
\end{figure}

Fig.~\ref{highfieldlines}a shows the \bnmr\ spectrum in the normal
state of \NS\ in a magnetic field of 300 mT applied along the c-axis
and beam direction, which are both perpendicular to the surface.
Above $T_c=7.0$ K the lineshape is nearly independent of magnetic
field, temperature and implantation depth. In a previous study we
showed that the Korringa relaxation in the normal state is anomalously
small compared to simple metals such as Ag \cite{Wang05}. Also, there
is no resolved quadrupolar splitting. The observed line width is
attributed to nuclear dipolar broadening from the $^{93}$Nb nuclear
moments. These results suggest that Li occupies a site in the van der
Waals gap between the \NS\ layers and is only weakly coupled to the
conduction band.

After field-cooling below $T_c$, the resonance broadens asymmetrically
(see Fig.\ref{highfieldlines}b) and exhibits the features
characteristic of a triangular vortex lattice in a superconductor. In
particular note the most probable frequency, or cusp frequency, shifts
by an amount $\Delta_c$ below the normal state frequency. The cusp
frequency (or magnetic field) arises from Li located midway between
two adjacent vortices. In addition, there is a high frequency tail
which corresponds to the magnetic field distribution near the vortex
core. The high frequency cutoff, which is shifted by an amount
$\Delta_v$ above the normal state frequency, corresponds to the field
at the vortex core. At high implantation energies we expect the
magnetic field distribution (frequency distribution) to be close to
that in a bulk superconductor. Such bulk field distributions have
been studied extensively with muon spin rotation (\msr\ ) and used to
extract the magnetic penetration depth and vortex core radius. In
applied magnetic fields ($H<<H_{c2}$) the local magnetic field at
position ${\bf r}=(x,y)$ relative to a vortex can be decomposed into
its Fourier components:
\begin{equation} \label{fourier}
B({\bf r})=B_0\sum_{\bf k} C(k,\rho)\frac{e^{-i{\bf k}\cdot {\bf r}}}{1+\lambda_{ab}^2k^2},
\end{equation}
where the sum is over all reciprocal lattice vectors of the vortex
lattice, $\lambda_{ab}$ is the in-plane penetration depth, $\rho$ is
the vortex core radius, $B_0$ is the average magnetic field, and
$C(k,\rho)$ is a phenomenological cutoff function which characterizes
the shape and size of the vortex core\cite{Brandt97}. $\rho$ is a
function of the coherence length, but may also depend on the
electronic structure of the vortex and vortex-vortex interactions.
The best fit\cite{fitcomment} of the current data is with a simple
Gaussian cutoff $C(k,\rho)=\exp{[-\frac{1}{2}k^2 \rho^2]}$ which gives
$\lambda_{ab}$=230(30) nm and $\rho$=13(1) nm. These parameters depend
slightly on the model for $C(k,\rho)$. For example using a Bessel
function cutoff \cite{Sonier00} gives $\lambda_{ab}=230(30)$ nm and
$\rho=11(1)$ nm. Previous \msr\ work on bulk \NS\ \cite{Sonier97}
obtained $\lambda_{ab} \approx 170$ and $\rho \approx 12$ at the same
field and temperature, using a Bessel function cutoff. The agreement
is reasonable considering there are substantial differences in the
experimental methods, the form of the raw data and resulting analysis.
For example, \bnmr\ spectra are acquired in the frequency domain
rather than in the time domain for \msr\ . Thus important features
such as $\Delta_v$ and $\Delta_c$ are evident in the raw \bnmr\ data
without data processing or analysis. Also, in \bnmr\ there is no
detectable background that interferes with the signal. On the other
hand the form of the field distribution is more complicated as
explained below.

\begin{figure}[h]
\includegraphics[width=8.0cm]{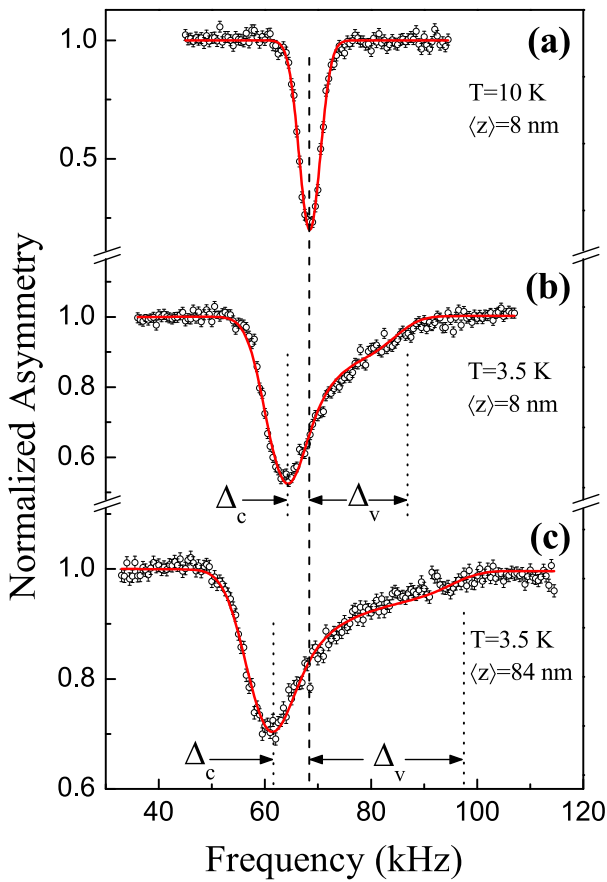}
\caption{(Color online) \bnmr\ spectra in \NS\ in a low magnetic field
  of $10.84$ mT. (a) In the normal state at $10$ K with a beam energy
  which corresponds to a mean implantation of $\langle z \rangle=8$
  nm. (b) The same conditions as (a) except field cooled to $0.5
  T_c$. (c) The same temperature and magnetic field as (b) but the
  mean implantation depth is about 10 times larger. }
\label{lowfieldlines}
\end{figure}

Fig.~\ref{lowfieldlines} shows \bnmr\ spectra measured in a lower
magnetic field of $10.8$ mT. The normal state resonance
(Fig.~\ref{lowfieldlines}a) is very similar to that observed at $300$
mT. The spectra in the superconducting state
(Figs.~\ref{lowfieldlines}b and \ref{lowfieldlines}c) show a strong
dependence on temperature (not shown) and implantation depth $\langle
z \rangle$. Several important features are evident from
Figs.~\ref{lowfieldlines}b and \ref{lowfieldlines}c without any
fitting. Firstly, $\Delta_c$ decreases as the mean depth changes from
$85$ nm (Fig.~\ref{lowfieldlines}c) to $8$ nm
(Fig.~\ref{lowfieldlines}b). Such narrowing of the field distribution
near the surface is understandable since the vortex lattice must
approach that of a bulk superconductor deep inside the material;
whereas, any line broadening from the vortex lattice must vanish well
outside the material. The actual crossover occurs when the mean depth
becomes comparable to $a_0/2\pi$ where $a_0 {\rm [nm]}
=1546/\sqrt{B_{ext} {\rm [mT]}}$ is the mean spacing between
vortices\cite{Niedermayer99}. Secondly, $\Delta_c$ in
Fig.~\ref{lowfieldlines}c has increased by about 50\% compared to the
corresponding lineshape at $300$ mT shown in
Fig.~\ref{highfieldlines}b. This is also consistent with the vortex
lattice model since the vortices are further apart in low field.

The most unexpected feature in the data is the remarkable similarity
between the lineshapes in Figs.~\ref{highfieldlines}b and
\ref{lowfieldlines}b, given the magnetic fields differ by a factor of
27. In particular, in Fig.~\ref{lowfieldlines}b note the cutoff at a
frequency $\Delta_v$ above the normal state frequency. Recall, this
corresponds to the frequency (magnetic field) at the vortex core. For
example using $\rho=13$ nm and $\lambda_{ab}=230$ nm one finds
$\Delta_v=76$ kHz or about three times larger than what is
observed. Typically $\Delta_v$ can only be observed in much higher
magnetic fields such as in Fig.~\ref{highfieldlines}b where the
vortices are close together. The small value of $\Delta_v$ in
Figs.~\ref{lowfieldlines}b and \ref{lowfieldlines}c indicates a very
extended vortex core where $\rho \gg \xi_{ab}$. The possible origin of
the over-sized vortices is discussed below.

\begin{figure}[h]
\includegraphics[width=8.0cm]{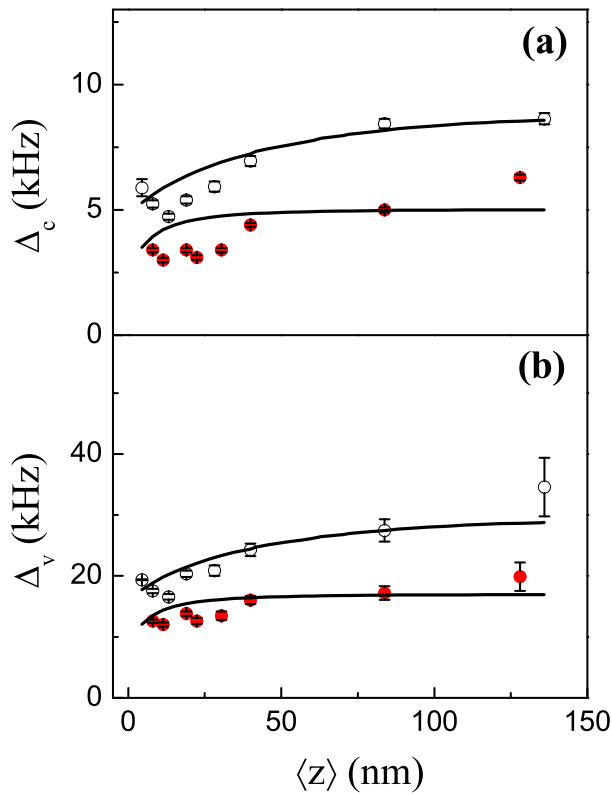}
\caption{(Color online) Depth dependence of the cusp frequency
$\Delta_c$ and vortex core frequency $\Delta_v$ relative to the normal
state frequency. The open circles are measured in a magnetic field of
10.8 mT while the filled circles are taken at 300 mT. In all cases
the temperature is 3.5K which corresponds to 0.5 T$_c$. The solid
curves are from the vortex lineshape model with a single depth
independent $\lambda_{ab}$ and $\rho$.}
\label{deltas}
\end{figure}

All the data in the superconducting state were fit to a vortex
lineshape model, which includes effects from the surface. Following
Ref.\cite{Niedermayer99}, we assume Laplace's equation is valid
outside the superconductor, and that a modified London model for a
triangular lattice of vortices applies inside the superconductor. This
yields a magnetic field at a depth $z$ equal to:
\begin{equation} \label{London}
B({\bf r},z)=B_0 \sum_{\bf k} \frac{C(k,\rho)}{\lambda_{ab}^2 \Lambda^2}\left[ 1-\frac{k}{\Lambda+k}e^{-\Lambda z} \right]e^{-i{\bf k}\cdot {\bf r}}
\end{equation}
where $\Lambda^2=k^2+1/\lambda_{ab}^2$. In analogy with bulk
superconductors we include a simple Gaussian cutoff function
$C(k,\rho)=\exp{[-\frac{1}{2}k^2\rho^2]}$. For each energy we
calculate a depth-averaged field distribution:
\begin{equation} \label{Listopping}
\langle n(B)\rangle =\int f(z) n(B,z) dz.
\end{equation}
where $f(z)$ is the stopping distribution obtained using the TRIM.SP
code to simulate the implantation of \Li\ in \NS\ \cite{trim} and
$n(B,z)$ is the field distribution at a well defined depth $z$. The
final step is to convolute the resulting frequency spectrum with a
Gaussian broadening function which takes into account the nuclear
dipolar fields and any residual disorder in the vortex
lattice\cite{Sonier00}. Typical fits to the model field distribution
are shown in Fig.~\ref{lowfieldlines}. The average frequency equals
the normal state frequency implying there is no measurable flux
expulsion, which is reasonable for our sample geometry. The measured
values for $\Delta_c$ and $\Delta_v$ are shown in Fig.~\ref{deltas}.
For comparison the solid lines are the best fit to the vortex
lineshape model assuming a depth independent $\lambda_{ab}$ and
$\rho$. At 10.8 mT we obtain $\lambda_{ab}=167(15)$ nm and
$\rho=77(10)$ nm compared to $\lambda_{ab}=230(30)$ nm and
$\rho=13(1)$ nm at 300 mT. At both fields $\Delta_c$ and $\Delta_v$
behave roughly according to this simple model, but there are clear
deviations. In particular $\Delta_c$ increases with decreasing
$\langle z \rangle$ close to the surface, whereas the simple model
predicts a monotonic decrease. Also, both $\Delta_c$ and $\Delta_v$
show a stronger variation with depth than the simple model
predicts. These are indications that the vortex interactions depend
slightly on depth.

The fitted value for $\lambda_{ab}$ at 10.8 mT is about 30\% smaller
than at 300 mT and is very close to the \msr\ results for bulk
NbSe$_2$. However, the most significant effect of the magnetic field
is the extremely large value of $\rho$ at 10.8mT. At 300 mT $\rho$ is
13(1) nm, which is close to the expected coherence length and
measurements of the core radius made with \msr\ in higher
fields\cite{Sonier97}. At $10.8$mT the effective $\rho$ increases to
77(10) nm. It should be stressed that changing the functional form of
the lineshape produces only small changes in $\rho$ and has no
influence on the ratio between $\rho$ in high and low field.
The magnitude of the effect is much larger than the calculated field dependence of the core radius originating  from the bound states which cause the Kramer-Pesch effect \cite{Ichioka99PRB}. It is interesting to compare these results with \msr\ in
bulk \NS. Early studies in this temperature range showed a substantial
increase in the vortex core size as the field was lowered but the
signal was not followed below 100 mT \cite{Sonier97}. The field
dependence we observe indicates a stronger variation with field below
100 mT than predicted from Ref. \cite{Sonier97} where the extrapolated
low field value for the core size was $25$ nm at $0.5 T_c$. Also, very
recent work at much lower temperatures (20 mK) gives an extrapolated
low field value of 10 nm \cite{Callaghan05}. A 60\% increase in $\rho$
below 400 mT is attributed to a multiband effect, whereby a second
smaller superconducting gap determines the vortex radius in low
fields. In a simple model, with minimal band coupling, the low field $
\xi_{ab} \approx v_F/\pi\Delta_0$ where $v_F$ is the Fermi velocity
and $\Delta_0$ is the smaller gap. We observe a much larger field
dependence to the core radius. The current measurement was made in a
lower magnetic field and of course much closer to the surface, but the
key difference is likely the temperature. Both ARPES \cite{Yokoya01}
and STM\cite{Rodrigo04} find evidence for a broad distribution of gaps
with parts of the Fermi surface showing no measurable gap at higher
temperatures near $T_c$. The ratio between the maximum and minimum
gaps is about five at $0.5T_c$ \cite{Rodrigo04}. Thermal vibrations of
the vortex lattice may also contribute to the observed field
dependence in the core radius, since in low magnetic fields the
vortices are further apart and less confined by their neighbors. The
upturn in $\Delta_c$ for small $\langle z\rangle $ (see
Fig.~\ref{deltas}) indicates some depth dependence to vortex
interactions as one approaches the surface. This is consistent with
vibrations since the vortex lattice is expected to stiffen very close
to the surface, either due to pinning or increased vortex repulsion
\cite{Pearl64}. A comprehensive theory of the vortices and their
interactions, which includes both electronic and vibrational
excitations is needed to resolve the precise origin of this effect.

In conclusion, low energy \bnmr\ has been used to measure the field
distribution of a vortex lattice below the surface of a
superconductor. Below $T_c$ in \NS\ the resonance line is
inhomogeneously broadened and has a lineshape which is characteristic
of a triangular vortex lattice. In a magnetic field of 300 mT the
magnetic penetration depth and core radius are similar to \msr\
measurements in bulk \NS. However, in a lower field of $10.8$ mT the
vortices are much larger. The multiband nature of the
superconductivity and thermal vibrations of the vortices are a
possible explanation.

We would like to thank Joe Brill at the University of Kentucky for
providing the sample. This research was supported by the Center for
Materials and Molecular Research at TRIUMF, the Natural Sciences and
Engineering Research Council of Canada and the Canadian Institute for
Advanced Research. We would especially like to acknowledge Rahim
Abasalti, Bassam Hitti, Donald Arseneau, and Suzannah Daviel for
expert technical support and Jeff Sonier for his critical reading of
the manuscript and helpful discussions.


\begin{thebibliography}{4}
\bibitem{fitcomment} The actual fit was done using Eq.~\ref{London} and  \ref{Listopping}, which takes into account the stopping distribution of Li and their proximity to the surface.
\end{thebibliography}
\end{document}